\documentclass[8pt,a4paper,twocolumn]{article}

\usepackage[switch]{lineno} 
\usepackage{multicol,lipsum}

\usepackage[]{cite}

\usepackage{subfiles}

\usepackage{abstract}

\usepackage[table, dvipsnames]{xcolor}
\definecolor{lightgray}{gray}{0.85}

\usepackage[top=1.8cm,left=1.7cm,right=1.7cm,bottom=1.75cm]{geometry}
\usepackage{relsize}
\usepackage{tabularx}
\usepackage{array}
\usepackage{wrapfig}
\usepackage{multirow}
\usepackage{booktabs}
\usepackage{caption}
\usepackage{graphicx}
\usepackage{xcolor}
\usepackage{comment}

\usepackage[affil-it]{authblk} 
\usepackage{etoolbox}

\usepackage{lmodern}
\usepackage{amsmath}
\usepackage{amsfonts}
\usepackage{mathtools}

\makeatletter
\patchcmd{\@maketitle}{\LARGE \@title}{\fontsize{16}{19.2}\selectfont\@title}{}{}
\makeatother

\begin{document}

\title{\textbf{{Polarization-insensitive wide-angle resonant acousto-optic phase modulator}}}

\author[1*]{Okan Atalar}
\author[1]{Amin Arbabian}

\affil[1]{\textit{Department of Electrical Engineering, Stanford University, Stanford, California 94305, USA}}
\affil[*]{Corresponding author: okan@stanford.edu\vspace{-2em}}

\date{}
\twocolumn[
  \begin{@twocolumnfalse}
\maketitle

\thispagestyle{empty}

\begin{abstract}
Phase modulators are commonly used devices in optics. Free-space phase modulators are typically constructed from optically anisotropic crystals exhibiting the Pockels effect. To preserve the light's polarization state as it propagates through the crystal, it is essential to align the polarization and angle of incidence of the light with respect to the crystal. In this study, we demonstrate the feasibility of constructing free-space resonant phase modulators with a broad acceptance angle and minimal dependence on the polarization state of light using an acousto-optic approach. These modulators operate in the megahertz frequency range, require modest power levels, have aperture sizes exceeding one square centimeter, and feature sub-millimeter thickness.\vspace{2em}
\end{abstract}

\end{@twocolumnfalse}
]

A phase modulator, as the name suggests, is a device used to manipulate the phase of an optical wave. Free-space phase modulators are commonly constructed by utilizing crystals that demonstrate the Pockels effect or by using liquid crystals. Liquid crystal modulators are efficient, compact, and mature, however, their operation bandwidth is limited to the kilohertz frequency regime~\cite{LC_1,LC_2}, limiting use cases.

Free-space electro-optic modulators generally operate at megahertz frequencies, have large apertures (to handle high-intensity laser beams and simplify alignment), and are commonly used in scientific settings~\cite{pockels_cell_primer,cw_laser_pockels_cell,recording_pockels_cell}. Electro-optic modulators configured for free-space operation are capable of achieving gigahertz modulation frequencies~\cite{plasmonic_phase_mod_free_space, gigahertz_mie_resonance_eo}, albeit with very limited apertures. Resonant designs are the preferred choice for applications that need a single frequency of operation, such as laser frequency stabilization~\cite{pound_drever_hall_frequency_stabilization} and mode locking~\cite{NdYAG_mode_locked,fm_and_am_mode_locking}, in order to minimize the required drive power.


Free-space resonant electro-optic phase modulators are constructed using a crystal showcasing the Pockels effect. Typically, optically anisotropic crystals such as BBO~\cite{bbo_pockels}, DKDP~\cite{DKDP_pockels}, and lithium niobate~\cite{eo_switcher_pockels_LN} are used due to their large electro-optic coupling coefficients and modest dielectric constants, translating to a high electro-optic figure of merit (a measure of the required drive power). The drawback for these modulators is the need to align both the polarization and the angle of incidence of the light with respect to the crystal. Although polarization insensitivity could be achieved by using custom crystal orientations (in exchange for higher operating power), optical birefringence causes the exit beam to have a different polarization state than the input beam for varying angles of incidence of light.

An alternative solution is to use optically isotropic materials that display the $\chi^{(2)}$ nonlinearity. Namely, materials belonging to the cubic crystal system with point group $\bar{4}3$m or $23$. Electro-optic phase modulators that are appropriately designed using these crystals do not require alignment of the polarization or incident angle of the incoming light. Some well known materials include gallium arsenide, gallium phosphide, cubic gallium nitride, cubic zinc selenide, and cubic zinc sulfide. The drawback of constructing modulators using such crystals is their low electro-optic figure of merit compared to optically anisotropic crystals~\cite{eo_cubic_crystals}. This results in a requirement for either excessively high power for operation or thicker crystals, as power consumption is inversely proportional to thickness.

Acousto-optic interactions enable various functionalities, including tunable filters~\cite{AOTF}, frequency shifters~\cite{frequency_shift_bragg_cell}, optical-frequency-comb generators~\cite{optical_frequency_comb_ao}, and high-speed spatial light modulators~\cite{doubly_resonant_ao}. Despite the diverse use cases, the existing acousto-optic modulation methods do not provide a solution for achieving polarization-insensitive and wide-angle phase modulation. Here, we build on our previous work constructing polarization modulators using longitudinal piezoelectric resonant photoelastic modulators~\cite{longitudinal_nat_paper,yz_ln_paper,optically_isotropic_paper}, and show that it is possible to construct resonant phase modulators that are polarization-insensitive and exhibit a wide acceptance angle. These modulators are made from optically isotropic and piezoelectric crystals, functioning at frequencies in the megahertz range, featuring input apertures on the order of square centimeters and thicknesses in the sub-millimeter range.


\begin{figure}[ht]
\centering
\includegraphics[width=\linewidth]{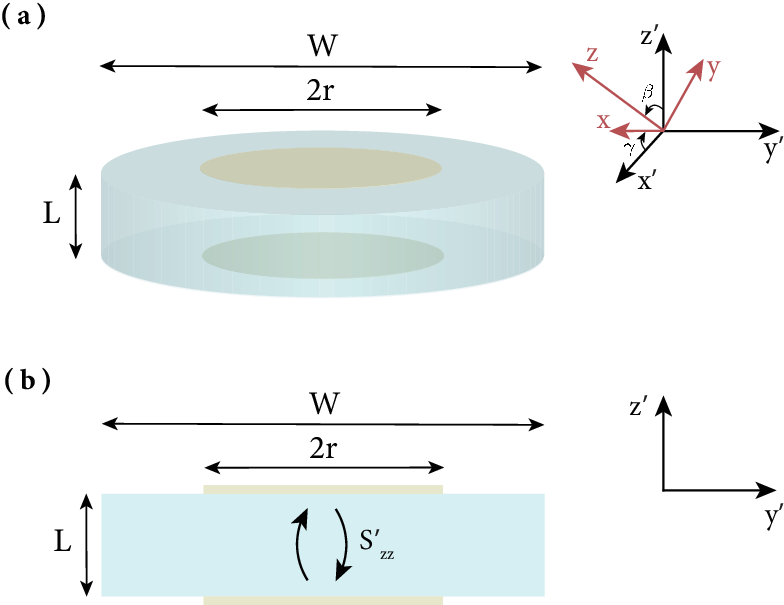}
\caption{Modulator design. (a) The modulator consists of a gallium arsenide wafer of thickness $L$, with a diameter of $W$ coated on top and bottom surfaces with transparent surface electrodes of radius $r$. The crystal orientation is shown with $(x,y,z)$ and the rotated coordinate system in $(x',y',z')$, where $z'$ direction is parallel to the wafer normal. The crystal coordinate system is obtained from the rotated system by applying rotations as shown, which correspond to ($\alpha = 0^{\circ}$, $\beta$, $\gamma$) rotations in Euler angles notation. (b) Side view of the modulator shown in (a). $S_{zz}'$ acoustic  standing wave is excited in the wafer by applying an RF signal to the surface electrodes that matches the resonance frequency for this strain.}
\label{fig:1}
\end{figure}

\begin{figure}[ht]
\centering
\includegraphics[width=\linewidth]{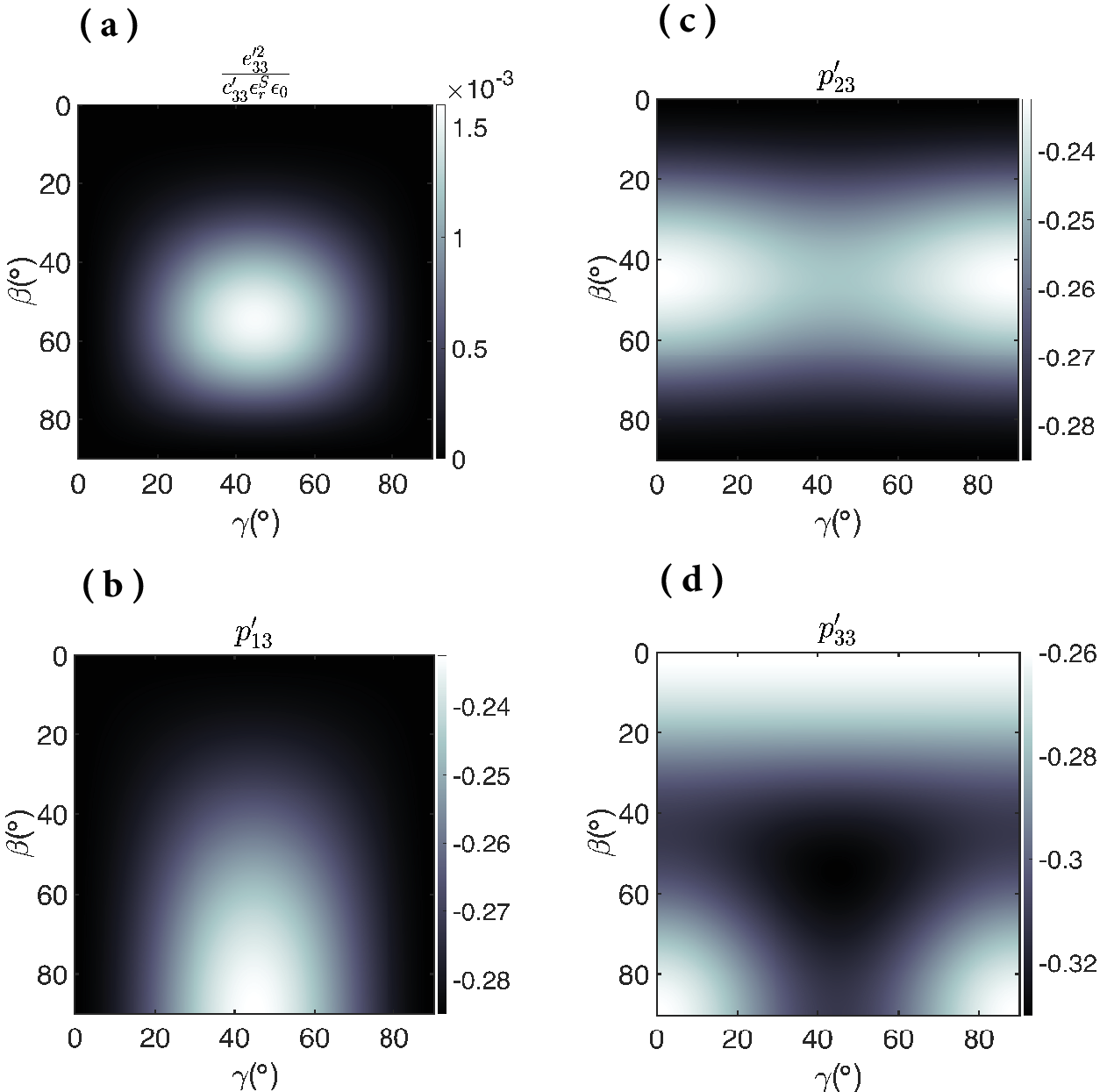}
\caption{The relevant coupling coefficients (in the rotated coordinate frame) corresponding to the $S'_{zz}$ longitudinal strain are plotted as a function of the crystal orientation. (a) The electromechanical coupling coefficient. (b) $p_{13}'$. (c) $p_{23}'$. (d) $p_{33}'$.}
\label{fig:2}
\end{figure}


\begin{figure}[ht]
\centering
\includegraphics[width=\linewidth]{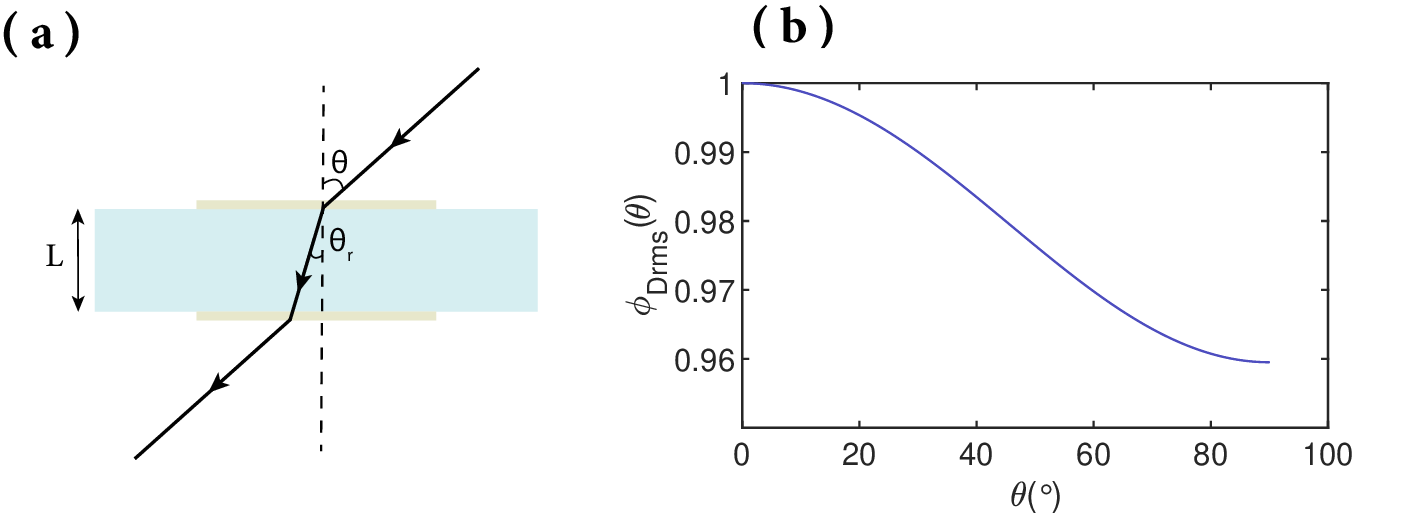}
\caption{Phase modulation amplitude as a function of the angle of incidence of light. (a) Refraction of a light beam incident on the modulator is shown. The incidence angle is $\theta$ with respect to the wafer normal and the refraction angle is $\theta_r = \text{sin}^{-1}\big(\frac{\text{sin}\theta}{n}\big)$. (b) Phase modulation amplitude ${\phi_{D}}_{rms}(\theta)$ plotted as a function of $\theta$.}
\label{fig:3}
\end{figure}

\begin{figure}[ht]
\centering
\includegraphics[width=\linewidth]{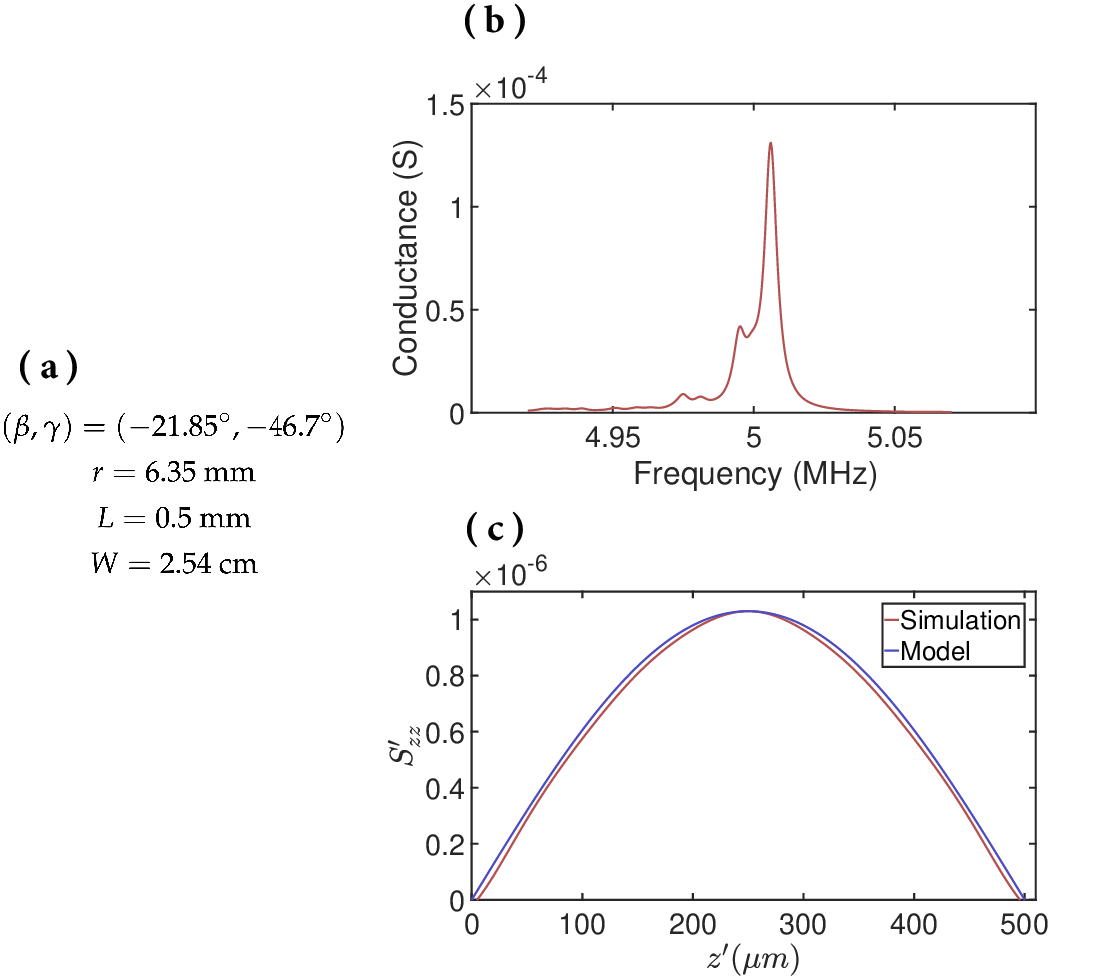}
\caption{Characteristics of the modulator. (a) The relevant parameters for the modulator are shown. (b) Conductance of the modulator is shown as a function of frequency. (c) The variation in $S_{zz}'$ parallel to the wafer normal and centered along the wafer is shown when the wafer is excited through the surface electrodes at 5.0061~MHz with 2Vpp.}
\label{fig:4}
\end{figure}

\begin{figure}[ht]
\centering
\includegraphics[width=\linewidth]{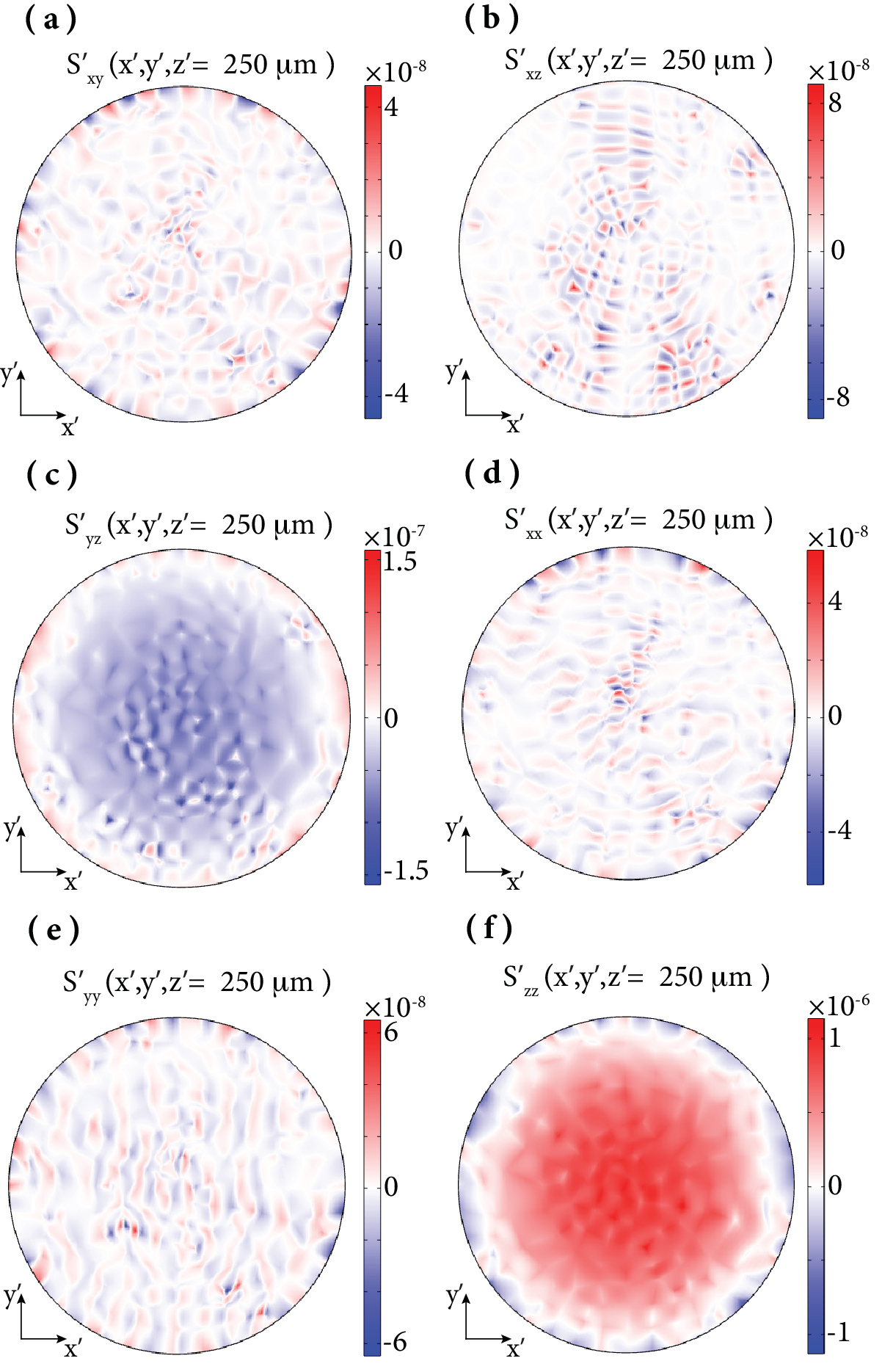}
\caption{Simulated strain amplitude profiles in the center of the wafer when the modulator is excited at 5.0061~MHz with 2Vpp applied to the surface electrodes. (a) $S'_{xy}$. (b) $S'_{xz}$. (c) $S'_{yz}$. (d) $S'_{xx}$. (e) $S'_{yy}$. (f) $S'_{zz}$.}
\label{fig:5}
\end{figure}

The design of the resonant acousto-optic phase modulator starts by choosing the material. In this work, we will show the design for GaAs, though other optically isotropic and piezoelectric materials such as GaP and cubic GaN could also be used. The next step is to choose the crystal dimensions. We choose to use a wafer to construct the modulator (similar to our previous work~\cite{optically_isotropic_paper}), that is coated on top and bottom surfaces with transparent surface electrodes. These electrodes serve as both the input aperture of the modulator as well as the excitation electrodes for the acoustic standing wave. The modulator dimensions and the relation between the rotated and crystal coordinate frame are shown in Fig.~\ref{fig:1}. The input aperture radius for the modulator is $r$, the modulator thickness is $L$, and the wafer diameter is $W$. The crystal orientation is shown with $(x,y,z)$ and the rotated coordinate system in $(x',y',z')$, where $z'$ direction is parallel to the wafer normal. $S_{zz}'$ acoustic standing wave is excited and confined to the electrode region of the modulator. In the primed tensor notation, $w'_{ij}$ denotes the element in the i\textsuperscript{th} row and j\textsuperscript{th} column of the rotated $\mathbf{w'}$ tensor (see Supplement 1 section 1 for more details). The resonance frequency ($f_r$) for the $S_{zz}'$ acoustic standing wave is expressed in Eq.~\eqref{Eq.1}, where $c_{33}'$ is the rotated stiffness coefficient, $e_{33}'$ is the rotated piezoelectric stress constant, $\epsilon_0$ is the vacuum permittivity, $\epsilon_r^S$ is the relative permittivity (evaluated at constant strain), and $\rho$ is the density of GaAs~\cite{auld_vol1}. For sub-millimeter thicknesses ($L < 1~\text{mm}$), the resonance frequency will be in the megahertz regime.

\begin{gather}
f_r \approx \frac{\sqrt{\big(c'_{33} + e'^2_{33}/(\epsilon_0 \epsilon_r^S)\big)/\rho}}{2L}.  \label{Eq.1}
\end{gather}

We next determine the crystal orientation. The crystal orientation influences both the piezoelectric coupling to the relevant strain as well as the photoelastic coupling coefficients pertaining to the excited acoustic standing wave (for this work, it is $S_{zz}'$). We would like to achieve efficient modulation of light (i.e., modulation with low drive power) as well as insensitivity to the polarization and angle of incidence of light. This is achieved if the relevant photoelastic coefficients pertaining to the relevant strain are large and equal ($p_{13}'$,$p_{23}'$,$p_{33}'$). The piezoelectric coupling coefficient to $S_{zz}'$ primarily influences the impedance matching condition. We try to avoid extremely small or excessively large coupling coefficients, as either extreme can constrain the quality factor ($Q$) of the modulator. The Butterworth-Van Dyke (BVD) equivalent resistance of the modulator is expressed in Eq.~\eqref{Eq.2} (derivation can be found in~\cite{yz_ln_paper}). The variation of the relevant piezoelectric and photoelastic coefficients as a function of the crystal orientation are plotted in Fig.~\ref{fig:2}.  

\begin{gather}
R_t \approx \frac{c'_{33}L}{2f_r \pi^2 r^2 e'^2_{33}Q}.  \label{Eq.2}
\end{gather}

Assuming that a crystal orientation can be found such that $p_{13}' = p_{23}' = p_{33}'$, the time-varying refractive index of GaAs is expressed as follows in Eq.~\eqref{Eq.3}, where $n$ is the refractive index of GaAs and the surface integral is carried out over the input aperture $A$ of the modulator with radius $r$. Here, we use the root-mean-square (rms) value since the acoustic standing wave will have spatial variation. The Pockels effect has been omitted in this analysis as the photoelastic effect is the dominant contributor.

\begin{gather}
\Delta n_{rms}(z') \approx  \frac{n^3p'_{13}}{2} \sqrt{\frac{\int_A S'^2_{zz}(x',y',z')dA}{\pi r^2}}. \label{Eq.3}
\end{gather}

The phase modulation amplitude for a wavelength of $\lambda_o$ is given as follows in Eq.~\eqref{Eq.4} for light incident at $(\theta,\psi)$ in spherical coordinates, where refraction of the light is accounted for. The wavevector direction of the incident light is expressed as $\hat{k} = \hat{a}'_x \text{sin}\theta \text{cos} \psi + \hat{a}'_y \text{sin}\theta \text{sin} \psi + \hat{a}'_z \text{cos}\theta$, where $\hat{a}_x'$, $\hat{a}_y'$, and $\hat{a}_z'$ are unit vectors corresponding to the directions $x'$, $y'$, and $z'$ of the rotated coordinate system, respectively. Since GaAs has a relatively large refractive index of approximately 3.55 at infrared wavelengths, the contribution of refraction is negligible. 

\begin{gather}
{\phi_{D}}_{rms}(\theta) \approx \frac{2 \pi \text{cos}\big(\text{sin}^{-1}(\frac{\text{sin}\theta}{n})\big)}{\lambda_o} \int_{0}^L \Delta n_{rms}(z') dz'. \label{Eq.4}
\end{gather}

Notice that the phase modulation amplitude is only a function of $\theta$, since we assumed that the time-varying birefringence of the modulator is independent of the polarization of the incident light ($p_{13}' = p_{23}' = p_{33}'$). The variation of the phase modulation amplitude as a function of $\theta$ is plotted in Fig.~\ref{fig:3}(b). Here, we assume that for perpendicular incidence of light ($\theta = 0^\circ$), the phase modulation amplitude equals 1 (this is an arbitrary choice, and any value could be used). For angle of incidence reaching $90^\circ$, we observe that the variation in the phase modulation amplitude is approximately 4\%. For incident light with electric field $\frac{\hat{p}}{2}E_oe^{j2\pi f_L t} + \text{c.c.}$, where $t$ stands for time, $f_L$ is the frequency of the incident light, $\hat{p}$ is the polarization direction of the incident light, and $\text{c.c.}$ stands for the complex conjugate, the electric field of the light at the output of the modulator is given as follows in Eq.~\eqref{Eq.5}, where $j$ is the imaginary unit with $j^2 = -1$ and $k = \frac{2 \pi n}{\lambda_o}$ the optical wavevector magnitude in GaAs. 

\begin{gather}
\bar{E}_f(\theta,t) = \frac{\hat{p}}{2} E_o \sum_{m=-\infty}^{\infty} j^m J_m\big({\phi_{D}}_{rms}(\theta)\big)e^{j\big((2\pi f_L + m2\pi f_r)t - kL\text{cos}\theta\big)} \nonumber \\ + \text{c.c.} \label{Eq.5}
\end{gather}

Generating strain in the modulator requires radio frequency (RF) power to be supplied to the surface electrodes. The RF power required to drive the modulator as a function of the strain generated in the active volume of the modulator is expressed as follows in Eq.~\eqref{Eq.6} (derivation can be found in~\cite{yz_ln_paper}), where the volume integral is carried out over the active volume $V = \pi r^2 L$ of the modulator.

\begin{gather}
P_{RF} \approx \frac{4 \pi f_{r} c'_{33} \int_V S'^2_{zz}(x',y',z')dV}{Q}. \label{Eq.6}
\end{gather}

It turns out that these two requirements, high efficiency and low-sensitivity to polarization and angle of incidence of light, can be met using crystal orientation $(\beta,\gamma) = (-21.85^\circ,-46.7^\circ)$ assuming we have a centimeter square scale input aperture, sub-millimeter thickness, and a $Q$ of approximately 1,000 (an estimate based on our previous work~\cite{optically_isotropic_paper}). For this crystal orientation, $(p_{13}',p_{23}',p_{33}') = (-0.2778,-0.2662,-0.2861)$ (see Supplement 1 section 1 for more details). Even though these coefficients are not equal, the maximum variation is approximately 7\%. Thus, the maximum variation of the phase modulation amplitude will be 7\% for any polarization of the incoming light. We note that this variation could be reduced by using other crystal orientations in exchange for higher operating power. Standard cuts, such as X-cut, Y-cut, and Z-cut also have a relatively small maximum variation of approximately 9\% for the relevant photoelastic coefficients. However, the relevant strain ($S_{zz}'$) cannot be excited since the relevant piezoelectric coupling coefficient ($e_{33}'$) is zero.


We next perform an electromechanical simulation of such a modulator using finite element solver COMSOL Multiphysics. We assume the following parameters, $(r,L,W,Q) = (6.35~\text{mm},0.5~\text{mm},2.54~\text{cm},1,000)$. The wafer with the longitudinal electrode geometry supports many acoustic modes, but we focus our attention in this work to the fundamental mode of the $S_{zz}'$ strain. This mode appears around 5~MHz, and the simulated conductance plot is shown in Fig.~\ref{fig:4}(b). The variation of the $S_{zz}'$ along the $z'$ direction is plotted in Fig.~\ref{fig:4}(c) when the modulator is driven at the peak of its conductance of $5.0061~\text{MHz}$. This variation should be sinusoidal (the model in the plot), since a fundamental acoustic standing wave is excited. The full simulated strain profiles of the wafer are shown in Fig.~\ref{fig:5} when the modulator is driven with 2Vpp at $5.0061~\text{MHz}$. Here, we see that a fairly uniform $S_{zz}'$ is excited, that is significantly larger in amplitude compared to the other strain components. The photoelastic coupling coefficients are also more than an order of magnitude larger for the $S_{zz}'$ strain compared to the shear strain components and comparable to the longitudinal strain components. We therefore approximate it as if only $S_{zz}'$ is excited in the wafer to simplify the analysis.

We compare our model to the simulation. Using Eq.~\eqref{Eq.1}, Eq.~\eqref{Eq.2}, and the rotated tensor parameters in Supplement 1 section 1, we estimate the resonant frequency and the BVD equivalent resistance pertaining to the $S_{zz}'$ mode as $4.96~\text{MHz}$ and $5.6~\text{k}\Omega$, respectively. From Fig.~\ref{fig:4}(b), we see that the resonant frequency from the simulation is $5.0061~\text{MHz}$, and the BVD equivalent resistance is approximately $7.6~\text{k}\Omega$. The discrepancy between theory and simulation for the BVD equivalent resistance is likely due to the non-uniformity of the strain generated in the modulator. 

We now calculate the required drive power to operate the modulator. A commonly used metric to evaluate power efficiency is the RF power required to attain a phase modulation amplitude of $\pi$ radians (i.e., ${\phi_{D}}_{rms}(\theta) = \pi$). For perpendicular incidence of light ($\theta = 0^\circ$) to the modulator and $\lambda_o = 940~\text{nm}$ (where GaAs is transparent), using Eq.~\eqref{Eq.3}, Eq.~\eqref{Eq.4}, Eq.~\eqref{Eq.5}, and the full tensor coefficients (see Supplement 1 section 1) for the chosen crystal orientation and modulator dimensions, we calculate this power as approximately 11.9~W. To make a comparison to a resonant electro-optic approach using GaAs with the same crystal thickness and input aperture, we find the power consumption as 617~W and 1.3~kW for the longitudinal and transverse interaction geometries, respectively (see Supplement 1 section 2). For the same crystal thickness and input aperture, the longitudinal acousto-optic approach requires significantly less RF power (greater than a factor of 50) to operate compared to the electro-optic approaches. The power consumption for the electro-optic approach could be reduced by using thicker crystals. To achieve the same power consumption as the acousto-optic approach, 2.6~cm and 5.5~cm thicknesses would be required for the longitudinal and transverse electro-optic modulators, respectively. These thicknesses would present a significant obstacle for various applications.

In summary, we have showcased the feasibility of achieving polarization-insensitive and wide-angle resonant phase modulation through a collinear acousto-optic method. These modulators are constructed from a piezoelectric crystal wafer that is optically isotropic. They feature centimeter square scale input apertures, sub-millimeter thicknesses, and operate at megahertz frequencies. The presence of a large input aperture and a wide acceptance angle would be advantageous when aligning a laser beam for the purpose of phase modulation. Polarization insensitivity allows for operation with unpolarized light, eliminating the need for polarizers and the associated loss of optical power. The proposed approach should function effectively across a wide optical bandwidth, spanning several hundred nanometers. This is achieved by utilizing anti-reflective coatings to minimize the residual amplitude modulation resulting from the cavity effect of the modulator. To enable operation in the visible wavelength range, other optically isotropic and piezoelectric materials, such as GaP, cubic GaN, cubic ZnSe, and cubic ZnS could be used to construct the modulator.

\section*{Funding.}
Stanford SystemX Alliance.

\section*{Acknowledgment.}
The authors thank Prof. Jonathan Simon for useful discussions.

\section*{Disclosures.}
The authors declare no conflicts of interest

\section*{Data availability.}
Data underlying the results presented in this paper are
not publicly available at this time but may be obtained from the authors upon
reasonable request.






\bibliographystyle{unsrt}
\bibliography{references}

\begin{thebibliography}{10}

\bibitem{LC_1}
Y.-C. Hsiao, C.-Y. Tang, and W.~Lee.
\newblock Fast-switching bistable cholesteric intensity modulator.
\newblock {\em Optics Express}, \textbf{19}(10):9744--9749, 2011.

\bibitem{LC_2}
J.~Sun, H.~Xianyu, Y.~Chen, and S.-T. Wu.
\newblock Submillisecond-response polymer network liquid crystal phase modulators at 1.06-$\mu$m wavelength.
\newblock {\em Applied Physics Letters}, \textbf{99}(2):021106, 2011.

\bibitem{pockels_cell_primer}
R.~Goldstein.
\newblock Pockels cell primer.
\newblock {\em Laser Focus}, 34(1), 1968.

\bibitem{cw_laser_pockels_cell}
R.P. De~Groote, I.~Budin{\v{c}}evi{\'c}, J.~Billowes, M.L. Bissell, T.E. Cocolios, G.J. Farooq-Smith, V.N. Fedosseev, K.T. Flanagan, S.~Franchoo, R.F.G. Ruiz, et~al.
\newblock {Use of a continuous wave laser and Pockels cell for sensitive high-resolution collinear resonance ionization spectroscopy}.
\newblock {\em Physical Review Letters}, 115(13):132501, 2015.

\bibitem{recording_pockels_cell}
D.N. Naik, G.~Pedrini, and W.~Osten.
\newblock {Recording of incoherent-object hologram as complex spatial coherence function using Sagnac radial shearing interferometer and a Pockels cell}.
\newblock {\em Optics Express}, 21(4):3990--3995, 2013.

\bibitem{plasmonic_phase_mod_free_space}
A.~Smolyaninov, A.~El~Amili, F.~Vallini, S.~Pappert, and Y.~Fainman.
\newblock Programmable plasmonic phase modulation of free-space wavefronts at gigahertz rates.
\newblock {\em Nature Photonics}, 13(6):431--435, 2019.

\bibitem{gigahertz_mie_resonance_eo}
I.-C. Benea-Chelmus, S.~Mason, M.L. Meretska, D.~L. Elder, D.~Kazakov, A.~Shams-Ansari, L.R. Dalton, and F.~Capasso.
\newblock Gigahertz free-space electro-optic modulators based on {Mie} resonances.
\newblock {\em Nature Communications}, 13(1):3170, 2022.

\bibitem{pound_drever_hall_frequency_stabilization}
E.D. Black.
\newblock {An introduction to Pound--Drever--Hall laser frequency stabilization}.
\newblock {\em American Journal of Physics}, 69(1):79--87, 2001.

\bibitem{NdYAG_mode_locked}
L.M. Osterink and J.D. Foster.
\newblock {A Mode-Locked Nd: YAG Laser}.
\newblock {\em Journal of Applied Physics}, 39(9):4163--4165, 1968.

\bibitem{fm_and_am_mode_locking}
D.J. Kuizenga and A.E. Siegman.
\newblock {FM} and {AM} mode locking of the homogeneous laser-{Part I}: {Theory}.
\newblock {\em IEEE Journal of Quantum Electronics}, 6(11):694--708, 1970.

\bibitem{bbo_pockels}
J.~Vengelis, G.~Sinkevi{\v{c}}ius, J.~Banys, L.~Masiulis, R.~Grigonis, J.~Domarkas, and V.~Sirutkaitis.
\newblock {Investigation of piezoelectric ringing effects in Pockels cells based on beta barium borate crystals}.
\newblock {\em Applied Optics}, 58(33):9240--9250, 2019.

\bibitem{DKDP_pockels}
A.~Starobor and O.~Palashov.
\newblock Thermal effects in the {DKDP Pockels} cells in the 215--300 {K} temperature range.
\newblock {\em Applied Optics}, 55(26):7365--7370, 2016.

\bibitem{eo_switcher_pockels_LN}
C.~Li.
\newblock {Electrooptic switcher based on dual transverse Pockels effect and lithium niobate crystal}.
\newblock {\em IEEE Photonics Technology Letters}, 29(24):2159--2162, 2017.

\bibitem{eo_cubic_crystals}
C.F. Buhrer, L.R. Bloom, and D.H. Baird.
\newblock Electro-optic light modulation with cubic crystals.
\newblock {\em Applied Optics}, 2(8):839--846, 1963.

\bibitem{AOTF}
S.E. Harris and R.W. Wallace.
\newblock Acousto-optic tunable filter.
\newblock {\em JOSA}, 59(6):744--747, 1969.

\bibitem{frequency_shift_bragg_cell}
J.B. Abbiss and W.T. Mayo.
\newblock Deviation-free {Bragg} cell frequency-shifting.
\newblock {\em Applied Optics}, 20(4):588--553, 1981.

\bibitem{optical_frequency_comb_ao}
S.N. Mantsevich, A.S. Voloshin, and K.B. Yushkov.
\newblock Optical-frequency-comb generation with collinear acousto-optic diffraction: Theory and simulations.
\newblock {\em Physical Review A}, 100(1):013829, 2019.

\bibitem{doubly_resonant_ao}
S.~Valle and K.C. Balram.
\newblock High-frequency, resonant acousto-optic modulators fabricated in a {MEMS} foundry platform.
\newblock {\em Optics Letters}, 44(15):3777--3780, 2019.

\bibitem{longitudinal_nat_paper}
O.~Atalar, R.~Van~Laer, A.H. Safavi-Naeini, and A.~Arbabian.
\newblock Longitudinal piezoelectric resonant photoelastic modulator for efficient intensity modulation at megahertz frequencies.
\newblock {\em Nature Communications}, 13(1):1526, 2022.

\bibitem{yz_ln_paper}
O.~Atalar, S.~Yee, A.H. Safavi-Naeini, and A.~Arbabian.
\newblock {YZ} cut lithium niobate longitudinal piezoelectric resonant photoelastic modulator.
\newblock {\em Optics Express}, 30(26):47103--47114, 2022.

\bibitem{optically_isotropic_paper}
O.~Atalar and A.~Arbabian.
\newblock Optically isotropic longitudinal piezoelectric resonant photoelastic modulator for wide angle polarization modulation at megahertz frequencies.
\newblock {\em JOSA A}, 40(12):2249--2258, 2023.

\bibitem{auld_vol1}
B.~A. Auld.
\newblock {\em Acoustic fields and waves in solids: volume I}.
\newblock Krieger publishing, 1973.

\bibitem{COMSOL5}
COMSOL Multiphysics version 5.0. https://www.comsol.com.

\bibitem{resonant_ao_coefficients}
P.~Renosi, J.~Sapriel, and B.~Djafari-Rouhani.
\newblock {Resonant acousto-optic effects in InP and GaAs and related devices}.
\newblock In {\em 1993 (5th) International Conference on Indium Phosphide and Related Materials}, pages 592--595. IEEE, 1993.

\bibitem{bond_matrices}
W.L. Bond.
\newblock The mathematics of the physical properties of crystals.
\newblock {\em The Bell System Technical Journal}, 22(1):1--72, 1943.

\end{thebibliography}


\onecolumn
\newpage


\begin{center}
  \section*{\textbf{\fontsize{16}{19.2}\selectfont Supplementary material}}
\end{center} 

\bigskip \bigskip 

\renewcommand\thefigure{S\arabic{figure}}
\setcounter{figure}{0} 

\setcounter{section}{0}


\section{Rotated tensors of GaAs}
The stiffness~\cite{auld_vol1}, piezoelectric~\cite{COMSOL5}, photoelastic~\cite{resonant_ao_coefficients}, and electro-optic tensors are shown below for GaAs in Voigt notation, where $\mathbf{c}$ is the stiffness tensor (evaluated at constant electric field), $\mathbf{e}$ is the piezoelectric stress tensor, $\mathbf{p}$ is the photoelastic tensor (for light wavelength of 940~nm), and $\mathbf{r}$ is the electro-optic tensor:

\begin{gather}
\mathbf{c} = 
\begin{pmatrix}
118.8 & 59.4 & 59.4 & 0 & 0 & 0  \\
59.4 & 118.8 & 59.4 & 0 & 0 & 0 \\
59.4 & 59.4 & 118.8 & 0 & 0 & 0 \\
0 & 0 & 0 & 53.8 & 0 & 0 \\
0 & 0 & 0 & 0 & 53.8 & 0 \\
0 & 0 & 0 & 0 & 0 & 53.8 \label{Eq.S1} \tag{S1}
\end{pmatrix} \text{(GPa)}
\end{gather}

\begin{gather}
\mathbf{e} = 
\begin{pmatrix}
0 & 0 & 0 & 0.14 & 0 & 0  \\
0 & 0 & 0 & 0 & 0.14 & 0 \\
0 & 0 & 0 & 0 & 0 & 0.14 \label{Eq.S2} \tag{S2}
\end{pmatrix} (\text{Cm}^{-2})
\end{gather}

\begin{gather}
\mathbf{p} = 
\begin{pmatrix}
-0.26 & -0.285 & -0.285 & 0 & 0 & 0  \\
-0.285 & -0.26 & -0.285 & 0 & 0 & 0 \\
-0.285 & -0.285 & -0.26 & 0 & 0 & 0 \\
0 & 0 & 0 & -0.04 & 0 & 0 \\
0 & 0 & 0 & 0 & -0.04 & 0 \\
0 & 0 & 0 & 0 & 0 & -0.04 \label{Eq.S3} \tag{S3} 
\end{pmatrix} 
\end{gather}

\begin{gather}
\mathbf{r} = 
\begin{pmatrix}
0 & 0 & 0 \\
0 & 0 & 0 \\
0 & 0 & 0 \\
1.7 & 0 & 0 \\
0 & 1.7 & 0 \\
0 & 0 & 1.7  \label{Eq.S4} \tag{S4} 
\end{pmatrix} (\text{pm/V})
\end{gather}

The rotated tensors in the primed coordinate notation are related to the crystal coordinate notation through the Bond transformation matrices~\cite{bond_matrices}. For Euler angles $(\alpha,\beta,\gamma)$, the relevant transformation matrices are expressed as follows: 

\begin{gather}
\mathbf{R} = 
\begin{pmatrix}
r_{xx} & r_{xy} & r_{xz} \\
r_{yx} & r_{yy} & r_{yz} \\
r_{zx} & r_{zy} & r_{zz} 
\end{pmatrix} =
\begin{pmatrix}
\text{cos}\alpha \text{cos}\gamma - \text{sin}\alpha \text{cos}\beta \text{sin} \gamma & -\text{cos}\alpha \text{sin}\gamma - \text{sin}\alpha \text{cos} \beta \text{cos}\gamma  & \text{sin}\beta \text{sin}\alpha \\
\text{sin}\alpha \text{cos} \gamma + \text{cos}\alpha \text{cos}  \beta \text{sin} \gamma & -\text{sin} \alpha \text{sin} \gamma + \text{cos}\alpha  \text{cos}\beta \text{cos}\gamma  & -\text{sin}\beta \text{cos} \alpha \\
\text{sin}\beta \text{sin}\gamma & \text{sin}\beta \text{cos}\gamma & \text{cos} \beta \label{Eq.S5} \tag{S5} 
\end{pmatrix}
\end{gather} 

\begin{gather}
\mathbf{M} = 
\begin{pmatrix}
r_{xx}^2 & r_{xy}^2 & r_{xz}^2 & 2r_{xy}r_{xz} & 2r_{xz}r_{xx} & 2r_{xx}r_{xy}  \\
r_{yx}^2 & r_{yy}^2 & r_{yz}^2 & 2r_{yy}r_{yz} & 2r_{yz}r_{yx} & 2r_{yx}r_{yy} \\
r_{zx}^2 & r_{zy}^2 & r_{zz}^2 & 2r_{zy}r_{zz} & 2r_{zz}r_{zx} & 2r_{zx}r_{zy} \\
r_{yx}r_{zx} & r_{yy}r_{zy} & r_{yz}r_{zz} & r_{yy}r_{zz} + r_{yz}r_{zy} & r_{yx}r_{zz} + r_{yz}r_{zx} & r_{yy}r_{zx} + r_{yx}r_{zy} \\
r_{zx}r_{xx} & r_{zy}r_{xy} & r_{zz}r_{xz} & r_{xy}r_{zz} + r_{xz}r_{zy} & r_{xz}r_{zx} + r_{xx}r_{zz} & r_{xx}r_{zy} + r_{xy}r_{zx} \\ 
r_{xx}r_{yx} & r_{xy}r_{yy} & r_{xz}r_{yz} & r_{xy}r_{yz} + r_{xz}r_{yy} & r_{xz}r_{yx} + r_{xx}r_{yz} & r_{xx}r_{yy} + r_{xy}r_{yx} \label{Eq.S6} \tag{S6} 
\end{pmatrix} 
\end{gather}

\begin{gather}
\mathbf{c'} = \mathbf{M}\mathbf{c}\mathbf{M}^T  \label{Eq.S7} \tag{S7} 
\end{gather}

\begin{gather}
\mathbf{e'} = \mathbf{R}\mathbf{e}\mathbf{M}^T  \label{Eq.S8} \tag{S8}
\end{gather}

\begin{gather}
\mathbf{p'} = \mathbf{M}\mathbf{p}\mathbf{M}^T  \label{Eq.S9} \tag{S9}
\end{gather}

\begin{gather}
\mathbf{r'} = \mathbf{M}\mathbf{r}\mathbf{R}^T \label{Eq.S10} \tag{S10}
\end{gather}

For Euler angles $(\alpha,\beta,\gamma) = (0^\circ,-21.85^\circ,-46.7^\circ)$, the rotated tensors are as follows:

\begin{gather}
\mathbf{c'} = \begin{pmatrix}
142.82 & 38.71 & 56.07 & 8.30 & 0.53 & -1.32  \\
38.71 & 148.13 & 50.76 & 4.89 & -0.46 & 1.14 \\
56.07 & 50.76 & 130.76 & -13.19 & -0.07 & 0.18 \\
8.3 & 4.89 & -13.19 & 45.16 & 0.18 & -0.46 \\
0.53 & -0.46 & -0.07 & 0.18 & 50.47 & 8.3 \\
-1.32 & 1.14 & 0.18 & -0.46 & 8.3 & 33.11
\end{pmatrix} \text{(GPa)} \label{Eq.S11} \tag{S11} 
\end{gather}

\begin{gather}
\mathbf{e'} = 
\begin{pmatrix}
0 & -0.0057 & 0.0057 & -0.0060 & 0.1297 & 0.0520  \\
0.0520 & -0.1344 & 0.0824 & -0.0758 & -0.0060 & -0.0057 \\
0.1297 & -0.0758 & -0.0539 & 0.0824 & 0.0057 & -0.0060 
\end{pmatrix} (\text{Cm}^{-2}) \label{Eq.S12} \tag{S12}
\end{gather}

\begin{gather}
\mathbf{p'} = 
\begin{pmatrix}
-0.3123 & -0.2399 & -0.2778 & -0.0181 & -0.0012 & 0.0029  \\
-0.2399 & -0.3239 & -0.2662 & -0.0107 & 0.0010 & -0.0025 \\
-0.2778 & -0.2662 & -0.2861 & 0.0287 & 0.0002 & -0.0004 \\
-0.0181 & -0.0107 & 0.0287 & -0.0212 & -0.0004 & 0.0010 \\
-0.0012 & 0.0010 & 0.0002 & -0.0004 & -0.0328 & -0.0181 \\
0.0029 & -0.0025 & -0.0004 & 0.0010 & -0.0181 &  0.0051
\end{pmatrix} \label{Eq.S13} \tag{S13}
\end{gather}


\section{RF Power Required for the Electro-Optic Approach}
\begin{figure*}[t!]
\centering
\includegraphics[width=1\textwidth]{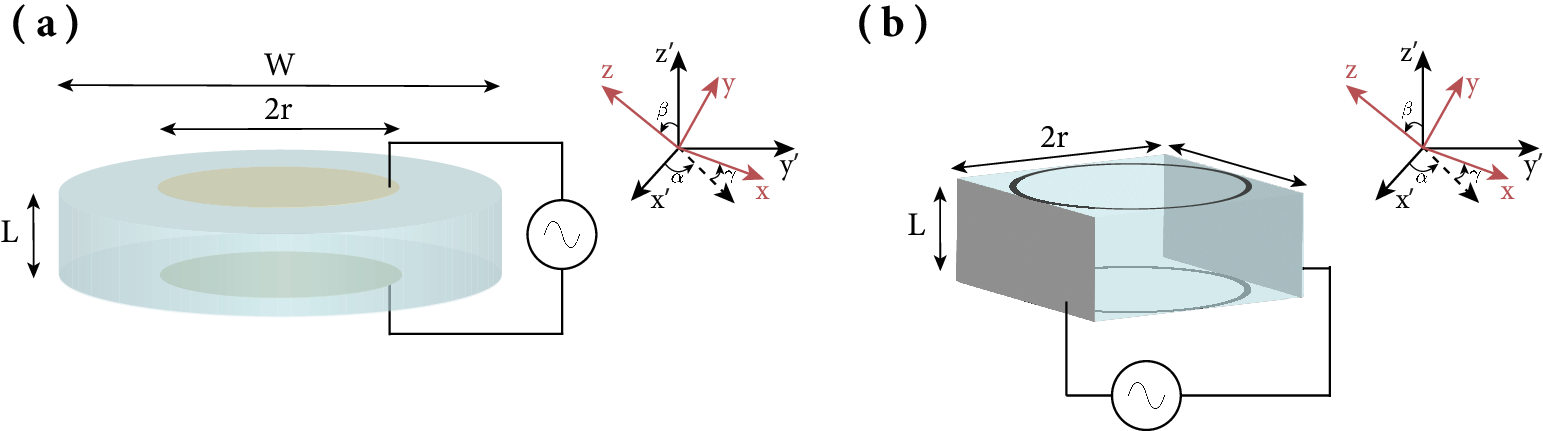}
\caption{Modulator schematics. (a) Longitudinal electro-optic modulator. The modulator consists of a GaAs wafer (with radius $r$ and thickness $L$) coated on top and bottom surfaces with transparent electrodes. (b) Transverse electro-optic modulator. The modulator consists of a GaAs cuboid (with dimensions $2r$ by $2r$ by $L$) that is coated on the sides with electrodes (does not have to be transparent). The input aperture is the inner region of the ring (with radius $r$).}
\label{fig:s1}
\end{figure*}

In this section, we calculate the RF power required to operate a free-space electro-optic modulator constructed using GaAs. Two different geometries could be used, transverse and longitudinal, and both will be calculated. The schematics of the modulator are shown in Fig.~\ref{fig:s1} for the longitudinal and transverse configurations. For both configurations, the crystal thickness is $L$ and the input aperture radius is $r$. The power consumption of the modulator is given below, where $C_M(\alpha, \beta, \gamma)$ is the capacitance, $V_M(\alpha, \beta, \gamma)$ is the amplitude of the applied sinusoidal RF voltage to the modulator electrodes, and $Q_{RF}$ is the RF quality factor (formed by connecting for instance a lumped inductor to the modulator).

\begin{gather}
P_{EO}(\alpha,\beta, \gamma) = \frac{C_M(\alpha, \beta, \gamma)V^2_M(\alpha,\beta, \gamma)f_r}{2Q_{RF}} \label{Eq.S14} \tag{S14}
\end{gather}

\subsection{Longitudinal Modulator}

For the longitudinal modulator, the capacitance is calculated as follows:

\begin{gather}
C_M(\alpha,\beta, \gamma) \approx \epsilon_0{\epsilon_r^S} \frac{\pi r^2}{L} \label{Eq.S15} \tag{S15}
\end{gather}

Here, we assume a single electro-optic coefficient $r'_{13}$ to calculate the minimum required power to drive the modulator. For crystal orientation $(\alpha,\beta,\gamma) = (72^\circ,235.8^\circ,225^\circ)$, $(r_{13}',r_{23}',r_{33}') = (1.001,0.96,-1.961)~\text{pm/V}$. This is the crystal orientation that maximizes $r_{13}'$ while minimizing the difference between $r_{13}'$ and $r_{23}'$ (this is more restrictive than also including $r_{33}'$, and so here we assume only perpendicular incidence of light to the modulator surface). With these assumptions, the amplitude of the phase modulation is expressed as follows:


\begin{gather}
{\phi_{D}}_{rms}(\theta = 0^\circ) \approx \frac{\pi n^3 V_M(\alpha,\beta, \gamma) r_{13}'}{\lambda_o} \label{Eq.S16} \tag{S16}
\end{gather}

We now calculate the half-wave voltage (voltage required to make the phase modulation amplitude equal to $\pi$ radians for this device):

\begin{gather}
V_{\pi}(\alpha,\beta, \gamma) = \frac{\lambda_o}{n^3 r_{13}'} \label{Eq.S17} \tag{S17}
\end{gather}

The RF power required to operate the modulator that corresponds to the half-wave voltage is expressed as follows:

\begin{gather}
{P_{EO}}_{\pi}(\alpha,\beta, \gamma) = \frac{\epsilon_0 \epsilon_r^S \pi r^2 f_r}{2LQ_{RF}} \Bigg(\frac{ \lambda_o}{n^3r_{13}'}\Bigg)^2 \label{Eq.S18} \tag{S18}
\end{gather}

We assume the following parameters to make a fair comparison to the acousto-optic approach: $L = 500~\mu \text{m}$, $f_r = 5~\text{MHz}$, $r = 6.35~\text{mm}$, $\lambda_o = 940~\text{nm}$, $n = 3.55$, $\epsilon_r^S = 12.5$, and $Q_{RF} = 50$. For these parameters, ${P_{EO}}_{\pi}(\alpha,\beta, \gamma) = 617~\text{W}$. Thus, 617~watts of RF power is required to induce $\pi$ radians amplitude of phase shift for these dimensions of the modulator. 


\subsection{Transverse Modulator}

For the transverse modulator, the capacitance is calculated as follows:

\begin{gather}
C_M \approx \epsilon_0{\epsilon_r^S} L \label{Eq.S19} \tag{S19}
\end{gather}

For the transverse geometry, we assume that the electric field is applied along $x'$ direction of the crystal (this is not a restrictive assumption, and any direction that is perpendicular to $z'$ would yield the same power consumption if the crystal orientation is optimized). Similar to the longitudinal modulator case, we assume a single electro-optic coefficient $r'_{11}$ to calculate the minimum required power to drive the modulator. For crystal orientation $(\alpha,\beta,\gamma) = (21.6^\circ,45^\circ,93.6^\circ)$, $(r_{11}',r_{21}',r_{31}') = (0.77,0.77,-1.54)~\text{pm/V}$. This is the crystal orientation that maximizes $r_{11}'$ while minimizing the difference between $r_{11}'$ and $r_{21}'$ (this is more restrictive than also including $r_{31}'$, and so here we assume only perpendicular incidence of light to the modulator surface). With these assumptions, the amplitude of the phase modulation is expressed as follows:

\begin{gather}
{\phi_{D}}_{rms}(\theta = 0^\circ) \approx \frac{\pi L n^3 r_{11}' V_M(\alpha,\beta, \gamma)}{2 r \lambda_o} \label{Eq.S20} \tag{S20}
\end{gather}

The half-wave voltage is calculated as follows:

\begin{gather}
V_{\pi}(\alpha,\beta, \gamma) = \frac{2 r \lambda_o}{L n^3 r_{11}'} \label{Eq.S21} \tag{S21}
\end{gather}

The RF power required to operate the modulator that corresponds to the half-wave voltage is expressed as follows:

\begin{gather}
{P_{EO}}_{\pi}(\alpha,\beta, \gamma) = \frac{\epsilon_0 \epsilon_r^S L f_r}{2Q_{RF}}\Bigg(\frac{2r\lambda_o}{Ln^3r_{11}'}\Bigg)^2 \label{Eq.S22} \tag{S22}
\end{gather}

Similar to the longitudinal modulator case, we assume the following parameters to make a fair comparison to the acousto-optic approach: $L = 500~\mu \text{m}$, $f_r = 5~\text{MHz}$, $r = 6.35~\text{mm}$, $\lambda_o = 940~\text{nm}$, $n = 3.55$, $\epsilon_r^S = 12.5$, and $Q_{RF} = 50$. For these parameters, ${P_{EO}}_{\pi}(\alpha,\beta, \gamma) = 1.3~\text{kW}$. Thus, 1.3~kilowatts of RF power is required to induce $\pi$ radians amplitude of phase shift for these dimensions of the modulator. 



\end{document}